\documentclass[letterpaper]{JHEP3}

\usepackage{epsfig}
\usepackage{graphicx}

\newcommand{\ie}{{\it i.e.,}\ }

\def\IR{{\hbox{{\rm I}\kern-.2em\hbox{\rm R}}}}
\newcommand{\be}{\begin{equation}}
\newcommand{\ee}{\end{equation}}
\newcommand{\beq}{\begin{equation}}
\newcommand{\eeq}{\end{equation}}
\newcommand{\beqa}{\begin{eqnarray}}
\newcommand{\eeqa}{\end{eqnarray}}

\title{{\huge The clash between de~Sitter \\ and anti-de~Sitter space}}

\author{Hassan Firouzjahi\footnote{E-mail: {\tt firouzh@hep.physics.mcgill.ca}} $\;$ and
$\,$ Fr\' ed\' eric Leblond\footnote{E-mail: {\tt
fleblond@hep.physics.mcgill.ca}}
\\
$^{*,\dagger}$ Department of Physics, McGill University, Montr\' eal, Qu\' ebec H3A 2T8\\
$^\dagger$ Perimeter Institute for Theoretical Physics, Waterloo,
Ontario N2J 2W9\\}

\abstract{We investigate solutions that are dynamically evolving
between asymptotically de~Sitter and asymptotically anti-de~Sitter
regions in the context of Einstein gravity coupled to general
matter fields in $d$ dimensions. We demonstrate the existence of a
{\it no go theorem} whenever the matter content of the theory is
`reasonable', \ie such that the weak energy condition is
satisfied. We show that there exist solutions for which the energy
conditions are violated in a finite region of the spacetime. We
speculate on the holographic interpretation of these gravitational
backgrounds by combining ideas from the AdS/CFT and the dS/CFT
dualities.}

\preprint{{\small hep-th/0209248}}

\begin{document}

\setcounter{footnote}{0}

\section{Introduction}

In this note we investigate spacetimes classically interpolating
between a region that has a positive cosmological constant and a
region with a negative cosmological constant. We find a {\it no go
theorem} when the matter fields present respect the weak energy
condition. Nevertheless, we demonstrate the existence of
gravitational solutions generated by matter fields violating the
energy conditions in a finite region of the spacetime. Such
solutions might, for example, be present in the Type II*
supergravity theories discussed in refs.~\cite{hull1,hull2}.

The paper is organized as follows: In section~\ref{asymptotia} we
describe the ansatz used for spacetimes with two asymptotic
regions: one including a de~Sitter (dS) boundary and the other
with an anti-de~Sitter (AdS) boundary. Then in
section~\ref{dynamics} we characterize the dynamics associated
with these gravitational backgrounds using the Einstein equations.
We derive a {\it no go theorem} when the matter fields coupled to
gravity violate the energy conditions. We close this section by
showing that there exists solutions when the weak energy condition
is relaxed and address how our results apply to asymptotically dS
and asymptotically AdS spacetimes. We conclude the presentation by
discussing how combining ideas from the
dS/CFT\cite{AS1,AS,BBM,witten} and the AdS/CFT\cite{adsrev}
dualities could have been interesting in the context of spacetimes
with both a dS and an AdS boundary.

\section{Spacetimes with dS and AdS asymptotia}
\label{asymptotia} Our starting point is a general metric ansatz
compatible with a spacetime dynamically evolving between two
regions with cosmological constants of different absolute value
and sign, \beq \label{metricgen} ds^{2} = - h(r) dt^{2} +
\frac{1}{h(r)} dr^{2} + a^{2}(r) \sum_{a} dx^{a}dx^{a}, \eeq where
$a=1,\, ... \, ,(d-2)$ and $d$ is the total number of dimensions.
The coordinates $x^{\mu}=(t,r,x^{a})$ all have the same range, \ie
$-\infty < x^{\mu} < +\infty$. To be more precise, we are
searching for spacetimes with two different asymptotic regions:
(1) one has a de~Sitter (dS) boundary, \ie  $I^{+}$ or $I^{-}$
(see ref.~\cite{AS3} for details), (2) the other includes a
boundary found in pure anti-de~Sitter (AdS) space (see
ref.~\cite{adsrev}). It is important at this point to stress that
we restrict ourselves to a family of solutions with these
characteristics. For example, this subset does not include all
spacetimes with both a region associated with a positive and a
negative cosmological constant. These do not necessarily include
dS and/or AdS boundaries.

The convention we use is that the coordinate $r$ is timelike
(spacelike) in the asymptotically dS (AdS) region. We choose the
boundary condition \beq \label{ds1} \lim_{r\rightarrow -\infty}
ds^{2} = \frac{r^{2}}{l^{2}_{\scriptscriptstyle dS}} dt^{2}
-\frac{l_{\scriptscriptstyle dS}^{2}}{r^{2}} dr^{2} +
\frac{r^{2}}{l_{\scriptscriptstyle dS}^{2}} d{\bf x}^{2}, \eeq
which corresponds to a spacetime with positive cosmological
constant, \beq \label{lds} \Lambda_{\scriptscriptstyle
dS}=\frac{(d-1)(d-2)}{2l_{\scriptscriptstyle dS}^{2}}. \eeq In
particular, by using the change of variables \beq \label{cov} r =
e^{r'/l_{\scriptscriptstyle AdS}}-e^{-r'/l_{\scriptscriptstyle
dS}}, \eeq the asymptotic expression (\ref{ds1}) takes the form
\beq \label{infla} -dr'^{2} + e^{-2r'/l_{\scriptscriptstyle
dS}}\left[ dt^{2} + d{\bf x}^{2} \right], \eeq which corresponds
to dS space in inflationary coordinates, where the flat slices
experience an exponential expansion. In this case the region
associated with the limit $r\rightarrow -\infty$ corresponds to
the spacelike boundary $I^{-}$. For $r\rightarrow +\infty$, we
impose the boundary condition \beq \label{ads1} \lim_{r\rightarrow
+\infty} ds^{2} = -\frac{r^{2}}{l_{\scriptscriptstyle AdS}^{2}}
dt^{2} +\frac{l_{\scriptscriptstyle AdS}^{2}}{r^{2}}dr^{2} +
\frac{r^{2}}{l_{\scriptscriptstyle AdS}^{2}} d{\bf x}^{2}, \eeq
which is the metric of a spacetime with a negative cosmological
constant, \beq \label{lads} \Lambda_{\scriptscriptstyle
AdS}=-\frac{(d-1)(d-2)} {2l_{\scriptscriptstyle AdS}^{2}}. \eeq
Using the change of coordinates (\ref{cov}) brings expression
(\ref{ads1}) to the form \beq \label{poincare} dr'^{2} +
e^{2r'/l_{\scriptscriptstyle AdS}}\left[ -dt^{2} + d{\bf
x}^{2}\right], \eeq which corresponds to the Poincar\'e patch of
AdS space with the boundary located at $r=+\infty$. In summary,
the boundary conditions we impose on the metric ansatz
(\ref{metricgen}) are \beq \label{bc1} \lim_{r\rightarrow \pm
\infty} a^{2}(r) = \frac{r^{2}}{l_{\scriptscriptstyle (A)dS}^{2}},
\;\;\;\;\;\; \lim_{r\rightarrow \pm \infty} h(r) = \pm a^{2}(r).
\eeq The results we present below are unaffected if we consider
the cases $r\rightarrow -r$ corresponding to the dS boundary being
located in the $r>0$ region and the AdS boundary in the $r<0$
region.

\section{Dynamics of the evolution}
\label{dynamics} In section \ref{asymptotia} we presented the
boundary conditions associated with a metric ansatz that has two
asymptotic regions: one which is time dependent and the other
which is static. In this section we present the details related to
the dynamics of such an evolution.

We begin by considering $d$-dimensional models of Einstein gravity
coupled to a real scalar field, \beq \label{action1} S =
\frac{1}{16\pi G_{\scriptscriptstyle N}} \int d^{d}x
\sqrt{-g}\left[ R
-(d-1)(d-2)g^{\mu\nu}\partial_{\mu}\phi\partial_{\nu}\phi-(d-1)(d-2)V(\phi)\right],
\eeq where $V(\phi)$ is a potential function for the scalar field.
We have normalized the scalar field terms in an unconventional
manner to simplify the equations of motion in the following
analysis. We use the convention $8\pi G_{{\scriptscriptstyle
N}}=1$ in what follows. Figure \ref{potential} illustrates a
typical potential that could lead to spacetimes that are
asymptotically dS and AdS. The equations of motion derived from
varying eq.~(\ref{action1}) are \beq \label{ein}
R_{\mu\nu}-\frac{1}{2}g_{\mu\nu}R=T_{\mu\nu}, \eeq where
$T_{\mu\nu}$ is the stress-energy tensor associated with the
scalar field, \beq \label{scalar} T_{\mu\nu} = (d-1)(d-2) \left[
\partial_{\mu}\phi \partial_{\nu}\phi - \frac{1}{2} g_{\mu\nu}
\left( g^{\lambda\rho}\partial_{\lambda}\phi\partial_{\rho}\phi +
V(\phi) \right) \right]. \eeq Considering solutions of the form
eq.~(\ref{metricgen}) with $\phi=\phi(r)$, we find that the
non-trivial components of the Einstein equations (\ref{ein}) are
\beq \label{ee1} (rr): \;\;\; \frac{d-2}{4}h' \frac{a^{2'}}{a^{2}}
+ \frac{(d-2)(d-3)}{8}h \left(\frac{a^{2'}}{a^{2}}\right)^{2} =
\frac{(d-1)(d-2)}{2}h\left[ (\phi')^{2} - \frac{V}{h} \right],
\eeq
\begin{eqnarray}
\label{ee2} (tt): \;\;\; \frac{d-2}{4}h'\frac{a^{2'}}{a^{2}} +
\frac{(d-2)(d-5)}{8}h \left(\frac{a^{2'}}{a^{2}}\right)^{2}
+\frac{d-2}{2}h \frac{a^{2''}}{a^{2}} \\ \nonumber =
-\frac{(d-1)(d-2)}{2}h \left[ (\phi')^{2} + \frac{V}{h}  \right],
\end{eqnarray}
\begin{eqnarray}
\label{ee3} (x^{a}x^{a}): \;\;\; \frac{h''}{2} + \frac{d-3}{2}
\left[ h' \frac{a^{2'}}{a^{2}} + h \frac{a^{2''}}{a^{2}} \right]
+\frac{(d-3)(d-6)}{8}h \left(\frac{a^{2'}}{a^{2}}\right)^{2} \\
\nonumber = -\frac{(d-1)(d-2)}{2}h \left[ (\phi')^{2} +
\frac{V}{h} \right],
\end{eqnarray}
where a `prime' denotes a derivative with respect to $r$. The
equation of motion for the scalar field, \beq \phi'' + \phi'
\left(\frac{h'}{h}+
\frac{d-2}{2}\frac{a^{2'}}{a^{2}}\right)-\frac{1}{2h}\frac{\partial
V }{\partial \phi} = 0,\eeq is automatically satisfied when the
Einstein equations are. Later on we will be interested in using a
version of eqs.~(\ref{ee1}), (\ref{ee2}) and (\ref{ee3}) where the
scalar field is replaced by a more general matter content. This is
simply accomplished by respectively replacing the RHS of these
equations by $T^{r}_{r}$, $T^{t}_{t}$ and $T^{x^{a}}_{x^{a}}$. It
is also useful to note that only two of these three equations of
motion are independent. \FIGURE{\epsfig{file=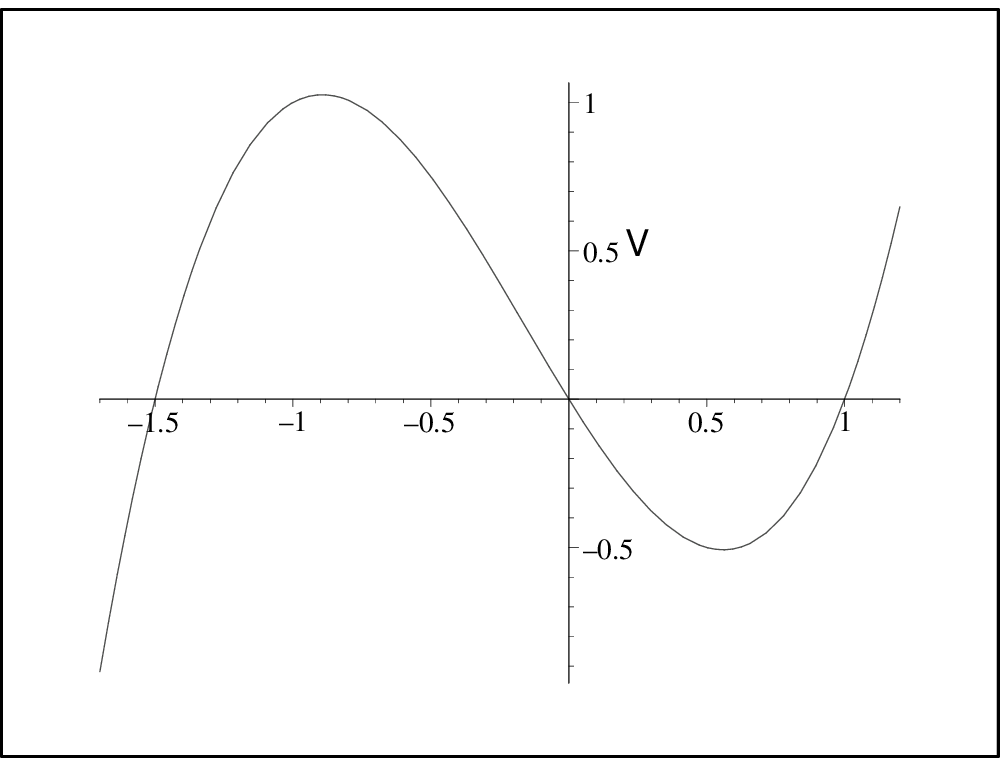,
width=6cm}\caption{A typical scalar field potential, $V(\phi)$,
for a transition from dS to AdS space. }\label{potential}}

Our strategy consists in using the Einstein equations in order to
verify whether or not there exist consistent solutions with
boundary conditions of the form (\ref{bc1}). Using a technique
introduced in ref.~\cite{CF}, we substract equations (\ref{ee2})
and (\ref{ee3}) which leads to \beq \left[ a^{(d-2)} \left( h' - h
\frac{a^{2'}}{a^{2}} \right) \right]'=0. \eeq This equation can be
written like \beq \label{replace} \left( a^{d} \left(
\frac{h}{a^{2}} \right)'\right)'=0, \eeq which has a solution of
the form \beq \label{the} h(r) = a^{2}(r)\left(k_{1} \int_{0}^{r}
\frac{dr'}{a(r')^{d}} + k_{2}\right), \eeq where $k_{1}$ and
$k_{2}$ are constants of integration. Imposing the dS/AdS boundary
conditions (\ref{bc1}) we obtain \beq k_{1}=2
\left(\int_{-\infty}^{+\infty} \frac{dr'}{a(r')^{d}}\right)^{-1},
\eeq \beq k_{2}=\frac{k_{1}}{2}\left(\int_{-\infty}^{0}
\frac{dr'}{a(r')^{d}} -\int_{0}^{\infty} \frac{dr'}{a(r')^{d}}
\right). \eeq We note that the case when the absolute value of the
cosmological constant is the same close to both boundaries
($l_{\scriptscriptstyle dS}=l_{\scriptscriptstyle AdS}$)
corresponds to $a(r)=\pm a(-r)$. For example, this implies
$k_{2}=0$ when the number of dimensions $d$ is even. Finally, by
subtracting eqs.~(\ref{ee1}) and (\ref{ee2}) we get \beq
\label{eom2} \frac{a''}{a} = -(d-1)(\phi')^{2}. \eeq In the
appendix we describe in more details the properties of a real
scalar field in a gravitational background of the form
(\ref{metricgen}).

Let us now write down eq.~(\ref{eom2}) for general matter fields,
\beq \label{geom2}
\frac{a''}{a}=\frac{T^{t}_{t}-T^{r}_{r}}{(d-2)h}. \eeq In the
region containing the boundary of AdS ($r>0$) we have $h(r)>0$.
The variable $t$ is then the timelike coordinate and
$T^{t}_{t}-T^{r}_{r}$ can be written like $-(\rho +p_{r})$ where
$\rho$ and $p_{r}$ are respectively the energy density and the
principal pressure along the coordinate $r$. In the region of
space with a dS boundary we have $h(r)<0$ and $r$ is a timelike
coordinate. This means that the combination $T^{t}_{t}-T^{r}_{r}$
is $(\rho +p_{t})$ where $p_{t}$ is the principal pressure along
the spacelike coordinate $t$. The weakest of the gravitational
energy conditions states (without proof) that a `reasonable'
gravitating system must be such that $T_{\mu\nu}n^{\mu}n^{\nu}\geq
0$ where $n^{\mu}$ is a null vector. In our case this means that
any region of the spacetime should have $\rho+p\geq 0$ where $p$
stands for all principal pressures taken individually. The weak
energy condition therefore requires that solutions associated with
a `reasonable' system have the property \beq \label{bcon}
\frac{a''}{a}=-\frac{\rho + p}{d-2}\leq 0, \eeq which restricts
considerably the number of possible candidate solutions for dS/AdS
spacetimes. \FIGURE{\epsfig{file=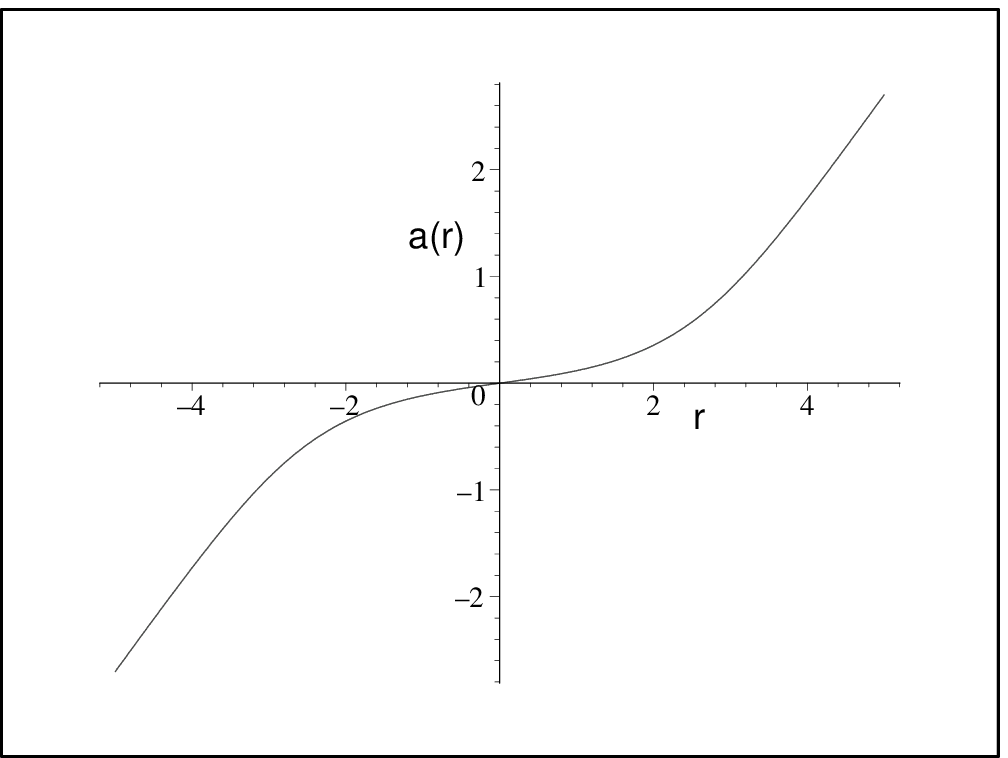,
width=6cm}\caption{Example of a function $a(r)$ with the boundary
conditions $a(r\rightarrow \pm\infty)=r/l_{(A)dS}$ where we have
set $l_{(A)dS}=1$. This curve has $a''>0$ for $r>0$ and $a''<0$
for $r<0$.} \label{f2}}

\section{No go theorem for dS/AdS spacetimes}
\label{criteria} In order to be acceptable, an asymptotically dS
and AdS (dS/AdS) solution must have the following three
properties: (1) its boundary conditions must be those specified in
eq.~(\ref{bc1}), (2) the weak energy condition (equivalent to
requiring $a''/a\leq 0$ everywhere) must be satisfied, and, (3) it
must not develop curvature singularities. We address these issues
in turn.

\subsection{Boundary behavior and weak energy condition}
There are four possible boundary conditions for dS/AdS solutions.
The first two possibilities correspond to \beq \label{bou1}
\lim_{r\rightarrow+\infty} a(r) = +\frac{r}{l_{\scriptscriptstyle
AdS}} \;\;\;\;\; {\rm and} \;\;\;\;\; \lim_{r\rightarrow -\infty}
a(r) = +\frac{r}{l_{\scriptscriptstyle dS}}, \eeq or, \beq
\label{bou2} \lim_{r\rightarrow+\infty} a(r) =
-\frac{r}{l_{\scriptscriptstyle AdS}} \;\;\;\;\; {\rm
and}\;\;\;\;\; \lim_{r\rightarrow -\infty} a(r) =
-\frac{r}{l_{\scriptscriptstyle dS}}. \eeq If
$l_{\scriptscriptstyle dS}=l_{\scriptscriptstyle AdS}$ these cases
correspond to $a(r)=-a(-r)$. Figures~\ref{f2} and \ref{f3} show
functions with the boundary behavior (\ref{bou1}). Functions such
as the one represented in figure~\ref{f2} are not solutions to the
equation of motion (\ref{bcon}) since they have $a''/a\geq 0$.
Functions of the type shown on figure~\ref{f3} are such that
$a''<0$ for $r>0$ and $a''>0$ for $r<0$ which means they are
acceptable candidate solutions. It is important to note that
functions with the boundary conditions (\ref{bou1}) and
(\ref{bou2}) always have at least one zero.

The remaining two boundary behaviors that can be considered
are\beq \label{bou3} \lim_{r\rightarrow +\infty} a(r) = +
\frac{r}{l_{\scriptscriptstyle AdS}} \;\;\;\;\;{\rm and}
\;\;\;\;\; \lim_{r\rightarrow -\infty} a(r) =
-\frac{r}{l_{\scriptscriptstyle dS}}, \eeq or \beq \label{bou4}
\lim_{r\rightarrow +\infty} a(r) = -\frac{r}{l_{\scriptscriptstyle
AdS}} \;\;\;\;\;{\rm and} \;\;\;\;\; \lim_{r\rightarrow -\infty}
a(r) = +\frac{r}{l_{\scriptscriptstyle dS}}. \eeq This corresponds
to $a(r)=a(-r)$ if we set $l_{\scriptscriptstyle
dS}=l_{\scriptscriptstyle AdS}$. Figure~\ref{f1} shows a function
with such boundary conditions. These functions always have at
least one minimum which implies that $a''/a>0$ at least in a small
region. This in turn means that eq.~(\ref{bcon}) is not satisfied.
Consequently, because they violate the weak energy condition,
solutions behaving like (\ref{bou3}) and (\ref{bou4}) are
rejected. \FIGURE{\epsfig{file=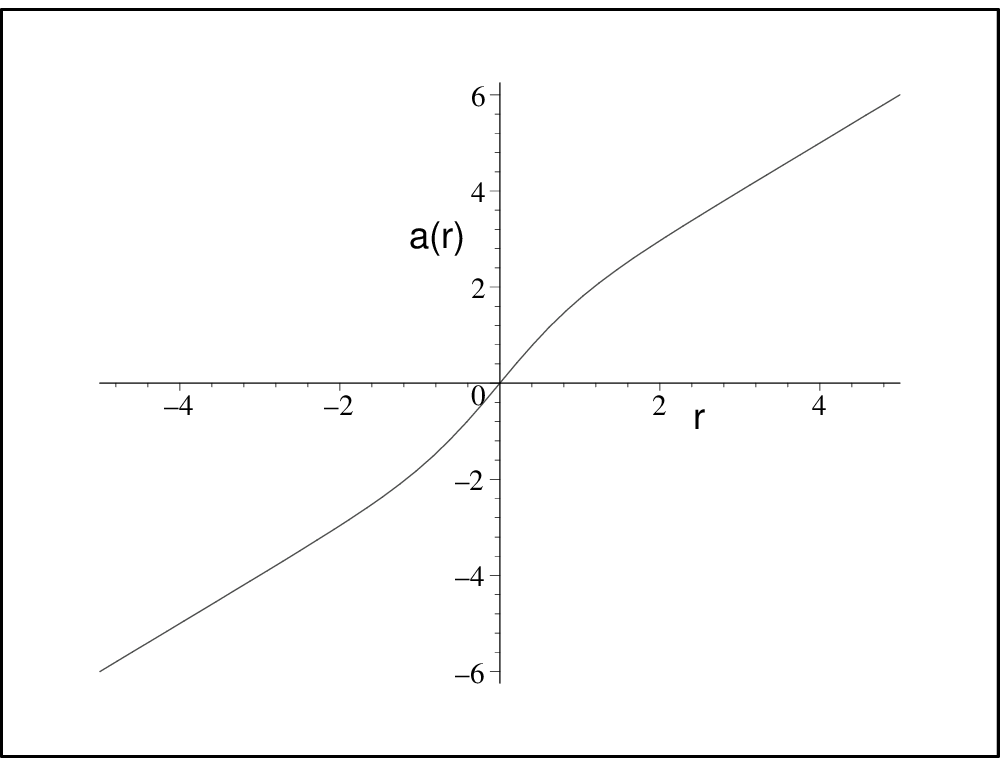,
width=6cm}\caption{Example of a function $a(r)$ with boundary
conditions $a(r\rightarrow -\infty)=r/l_{\scriptscriptstyle dS}$
and $a(r\rightarrow +\infty)=r/l_{\scriptscriptstyle AdS}$ where
we have set $l_{\scriptscriptstyle dS}=1=l_{\scriptscriptstyle
AdS}$. This example has $a''/a\leq 0$ everywhere.} \label{f3}}

\subsection{Curvature singularities}
We have found that there exists solutions $a(r)$ with the boundary
conditions (\ref{bc1}) that are not violating the energy
conditions. The function $a(r)$ then has an asymptotic behavior
given either by eq.~(\ref{bou1}) or eq.~(\ref{bou2}). These
solutions typically have a zero for some finite value of $r$. We
show that this leads to a curvature singularity. In fact, just by
inspection of eq.~(\ref{the}) it seems that $h(r)$ will diverge
when $a(r)=0$. The presence of curvature singularities can be
investigated by studying the behavior of the curvature invariants
$R$, $R_{\mu\nu}R^{\mu\nu}$ and
$R_{\mu\nu\rho\sigma}R^{\mu\nu\rho\sigma}$. These typically
involve the following terms, \beq \label{terms} h\frac{a''}{a}, \;
h'\frac{a'}{a}, \; h'', \; h\left(\frac{a'}{a}\right)^{2}. \eeq
From (\ref{terms}) it appears that whenever $h(r)$ and/or its
derivatives diverge a curvature curvature singularity will appear.
To illustrate how $h(r)$ itself diverges where $a(r)=0$, we
consider a subset of solutions for which $a(r)=-a(-r)$. We set
$l_{\scriptscriptstyle dS}=1=l_{\scriptscriptstyle AdS}$ to
simplify the analysis. These solutions must have the property,
\beq \label{close} \lim_{r\rightarrow 0^{\pm}} a(r) = a_{n}r^{n}
\;\;\;\;\; {\rm no\; sum},\eeq where $n$ is an integer and
$a_{0}=0$. Using eq.~(\ref{the}) we obtain \beq \lim_{r\rightarrow
0^{\pm}}h(r) = \frac{k_{1}}{1-nd}r^{1-n(d-2)}+k_{2}r^{2n}.\eeq
Finiteness of this expression requires that \beq 0\leq n \leq
\frac{1}{d-2}, \eeq which is only satisfied for $n=1$ and $d=3$.
We conclude that for $d>3$ the function $h(r)$ always diverges
when $a(r)=0$. Next we consider how $a'/a$ behaves when $a(r)=0$.
Using (\ref{close}) we find \beq \lim_{r\rightarrow 0^{\pm}}
\frac{a'}{a} = \frac{n}{r}, \eeq which clearly diverges. This
divergence of $a'/a$ is quite generic for all $d$ so the curvature
invariants for $d=3$ will diverge as well.

We showed that solutions with the property $a(r)=-a(-r)$
generically develop a curvature singularity at $r=0$. This also
applies to more general solutions for which $l_{\scriptscriptstyle
dS}\neq l_{\scriptscriptstyle AdS}$. In other words this statement
is true for all solutions with boundary conditions of the form
(\ref{bou1}) or (\ref{bou2}). We note for future reference that
solutions of the form shown in figure~\ref{f1} with boundary
conditions (\ref{bou3}) are perfectly well-behaved. This statement
can be traced to the fact that $a(r)$ does not have zeroes in this
case. It is unfortunate that these functions are not solutions to
the Einstein equations when the matter content of the theory
respects the weak energy condition.

The results of this section allows us to formulate the following
simple theorem:
\newenvironment{descbf}[1]{\begin{quote}{\em #1\/}:}{\end{quote}}
\begin{descbf}{{\bf No go theorem for dS/AdS transitions}}
Einstein gravity coupled to matter obeying the {\it weak energy
condition} has no solutions corresponding to a dynamical system
evolving from an asymptotic region containing a de~Sitter
(anti-de~Sitter) boundary to another region that includes an
anti-de~Sitter (de~Sitter) boundary.
\end{descbf} Of course this possibly excludes spacetimes without boundaries and
most certainly does not address spacetimes for which the evolution
is quantum mechanically driven. For example, instanton solutions
easily evade the {\it no go theorem} since they typically violate
the energy conditions.

\subsection{Regular solutions violating the energy conditions}
\label{wec}

The weak energy condition proposes that all physical systems have
$\rho + p \geq 0$. We found that solutions $a(r)$ to
eq.~(\ref{geom2}) with the boundary conditions (\ref{bc1}) and for
which the weak energy condition is satisfied do not exist. We also
found that if the weak energy condition is relaxed, regular
solutions are easily obtained. For example, the solution
illustrated on figure \ref{f1} does not lead to any curvature
singularity. Given such a solution, $h(r)$ can easily be obtained
using eq.~(\ref{the}) which gives a full characterization of the
dS/AdS spacetime.

We therefore conclude that there exist regular dS/AdS solutions
only when energy conditions are violated. Interesting examples of
such spacetimes are found in theories containing scalar fields
with a kinetic term of the wrong sign ($\phi\rightarrow i \phi$ in
(\ref{action1})). In fact, the equation of motion (\ref{eom2})
then becomes \beq \label{eo} \frac{a''}{a} = (d-1)(\phi')^{2},
\eeq which, as shown above, does have regular solutions. The Type
II* supergravity theories introduced in ref.~\cite{hull1,hull2}
generically contain fields with kinetic terms of the wrong
sign\footnote{See, for example, ref.~\cite{malik} where the Type
IIb* theory is used to construct elliptic dS space and find its
SO($N$) Euclidean super Yang-Mills holographic dual.}. These
theories are therefore natural laboratories where dS/AdS solutions
might be investigated further.
\FIGURE{\epsfig{file=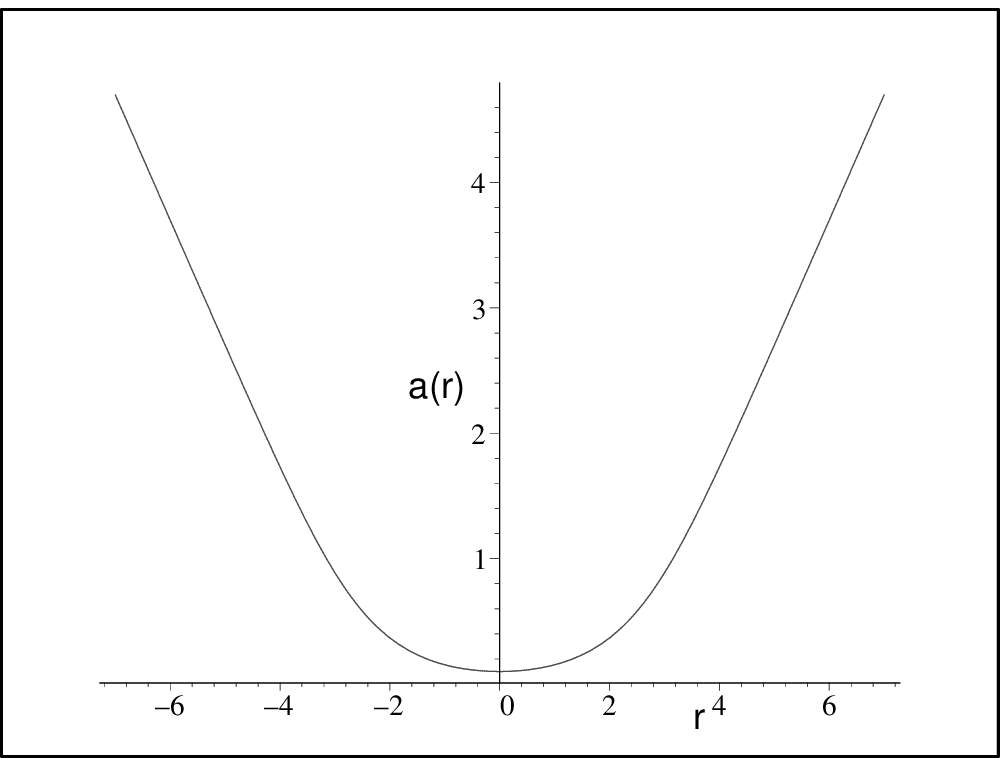, width=6cm}\caption{Example of
a function $a(r)$ with boundary conditions $a(r\rightarrow
-\infty)=-r/l_{\scriptscriptstyle dS}$ and $a(r\rightarrow
+\infty)=r/l_{\scriptscriptstyle AdS}$ where we have set
$l_{\scriptscriptstyle dS}=1=l_{\scriptscriptstyle AdS}$. This
curve has $a''/a\geq 0$ for all values of $r$.} \label{f1}}

\subsection{Asymptotically (A)dS geometries} \label{homo}
We generalize our analysis to include asymptotically dS (dS/dS)
and asymptotically AdS (AdS/AdS) spacetimes. The metric ansatz
(\ref{metricgen}) can be used to describe asymptotically dS
spacetimes when $h(r)=-a^{2}(r)$, \beq \label{dsds} ds^{2} =
-\frac{1}{a^{2}(r)}dr^{2} + a^{2}(r) \left(dt^{2}+d{\bf
x}^{2}\right). \eeq Using a change of coordinates of the form \beq
\label{changev} dr^{2}/a^{2}(r)=d\rho^{2}\eeq takes the metric to
a more familiar expression, \ie  \beq \label{expods} ds^{2} =
-d\rho^{2} + a^{2}(\rho) \left( dt^{2} + d{\bf x}^{2} \right) \eeq
which corresponds to a foliation with flat constant $\rho$
spacelike slices. Similarly, metrics suitable for the study of
AdS/AdS spacetimes are obtained by setting $h(r)=a^{2}(r)$ in
(\ref{metricgen}) which leads to \beq \label{adsads} ds^{2} =
\frac{1}{a^{2}(r)}dr^{2} + a^{2}(r) \left(-dt^{2}+d{\bf
x}^{2}\right). \eeq Using the change of coordinates
(\ref{changev}) this metric becomes \beq\label{expoads} ds^{2} =
d\rho^{2} + a^{2}(\rho) \left( -dt^{2} + d{\bf x}^{2} \right),
\eeq which corresponds to a foliation with flat Lorentzian slices.
When considering metric ansatz (\ref{dsds}) and (\ref{adsads}),
the boundary conditions corresponding to asymptotically (A)dS
geometries are \beq \label{bcfinal} \lim_{r\rightarrow \pm \infty}
a^{2}(r) = \frac{r^{2}}{l^{2}_{(A)dS}}. \eeq This last condition
is equivalent to requiring that there are two (A)dS boundaries in
the spacetime.

The boundary conditions (\ref{bcfinal}) are the same as those used
for dS/AdS spacetimes. The analysis we performed in
section~\ref{criteria} can therefore easily be extended to
(A)dS/(A)dS geometries. In fact, a certain combination of the
Einstein equations again leads to the constraint \beq
\frac{a''}{a} = -\frac{\rho + p}{(d-2)}, \eeq where the weak
energy condition dictates that $\rho + p \geq 0$. We have shown in
section~\ref{criteria} that there are no regular solutions $a(r)$
with boundary conditions (\ref{bcfinal}) respecting the weak
energy condition. The only exception in this case is $a(r)=r^2$
which corresponds to pure (A)dS space. We comment on that special
case at the end of this section. Our result for (A)dS/(A)dS
spacetimes does not imply that such geometries do not exist. The
correct statement is that there are no (A)dS/(A)dS spacetimes for
which two boundaries can be seen when the metric ansatz assumes a
foliation by flat ($d-1$)-dimensional hypersurfaces.

A generating technique for AdS/AdS spacetimes was introduced in
ref.~\cite{gubser} and generalized to dS/dS geometries in
ref.~\cite{tale1}. We now show that solutions such as those
typically only include one boundary which means that our metric
ansatz (\ref{metricgen}) with boundary conditions (\ref{bc1})
excludes them. For metrics of the form (\ref{expods}) and
(\ref{expoads}) one of the Friedmann equations can be written
like, \beq \ddot{(\ln a)} = -\frac{\rho +p}{d-2}, \eeq where a
`dot' signifies a derivative with respect to the variable $\rho$.
The only type of function which solves this equation is $\ln a$ of
the form represented on figure~\ref{f3}. This corresponds to a
scale factor with the boundary conditions, \beq
\lim_{\rho\rightarrow +\infty} a(\rho) = e^{\rho/l^{f}_{(A)dS}},
\;\;\;\;\; \lim_{\rho\rightarrow -\infty} a(\rho) =
e^{\rho/l^{i}_{(A)dS}}, \eeq where the upper indices $i$ and $f$
respectively label the length scales as $r\rightarrow -\infty$ and
$r\rightarrow +\infty$. Therefore we find that the scale factor
vanishes for $r=-\infty$ which means that this region corresponds
to an horizon, not a boundary. These solutions are regular (no
curvature singularity) asymptotically (A)dS spacetimes. It is easy
to demonstrate that these spacetimes cannot be written in the form
(\ref{dsds}) and (\ref{adsads}). Using the appropriate change of
coordinates we get, \beq r(\rho) = r(0) + \int_{0}^{\rho} a(\rho')
d\rho', \eeq but since $a(\rho\rightarrow -\infty)=0$ we see that
the domain of $r$ is limited to positive values while the metric
ansatz (\ref{dsds}) and (\ref{adsads}), in order to include two
boundaries, had to be such that $-\infty<r<+\infty$.

\paragraph{A global metric for pure dS space:}
We have indicated previously that the metric \beq \label{global} -
\frac{l^{2}}{r^{2}}dr^{2} + \frac{r^{2}}{l^{2}} \left( dt^{2} +
d{\bf x}^{2} \right), \eeq corresponds to a solution to the
Einstein equations with a cosmological constant, \beq \Lambda =
\frac{(d-1)(d-2)}{2l^{2}}. \eeq This solution corresponds to a
foliation of dS space with flat constant $r$ hypersurfaces.
Eq.~(\ref{global}) for dS is global in the sense that it covers
the whole spacetime. This can be seen by using the change of
coordinates $r=\pm l\, e^{\pm \tau/l}$. The original metric then
becomes \beq \label{inflag} ds^{2} = -\frac{1}{l^{2}}d\tau^{2} +
e^{\pm 2\tau/l}\left(dt^{2} + d{\bf x}^{2} \right), \eeq where
$-\infty<\tau<+\infty$. Expressions (\ref{inflag}) represent the
two inflationary patches of dS space. The Big Bang patch
(containing the boundary $I^{+}$ and corresponding to the plus
sign in (\ref{inflag})) is covered by the system of coordinates
(\ref{global}) for $-\infty<r<0$. The Big Crunch patch of dS
(containing the boundary $I^{-}$ and corresponding to the minus
sign in (\ref{inflag})) is covered by (\ref{global}) when
$0<r<+\infty$. Therefore the coordinate system (\ref{global}) for
pure dS covers the whole spacetime with flat ($d-1$)-dimensional
hypersurfaces.

\section{Discussion}
We used Einstein gravity coupled to general matter fields in order
to prove that when the weak energy condition is respected there
cannot be non-singular solutions which extrapolate between an
asymptotically de~Sitter region containing a boundary and an
asymptotically anti-de~Sitter region containing a boundary. This
result is quite generic. For example, there is no dS/AdS solutions
in any of the conventional supergravity theories. Exceptions might
appear when one considers sources in the form of such stringy
objects as orientifold planes and negative tension branes which
typically violate the weak energy condition (see, for example,
\cite{marolf}).

We conclude this presentation by discussing ideas from both the
AdS/CFT and the dS/CFT in order to initiate a discussion as to how
dS/AdS spacetimes might be interpreted in the context of
gauge/gravity dualities. In the AdS/CFT correspondence, quantum
gravity in pure AdS space is conjectured to be dual to a conformal
field theory in one less dimension that can be regarded as being
defined on the boundary of the bulk spacetime. For AdS in
Poincar\'e coordinates, \beq \label{poin2} ds^{2}=dr^{2} +
e^{2r/l_{\scriptscriptstyle AdS}}\left( -dt^{2} + d{\bf
x}^{2}\right), \eeq the boundary is located at $r=+\infty$. The
dS/CFT correspondence analogously proposes that quantum gravity in
dS space has a Euclidean conformal field theory dual defined on
its boundary(ies). When formulated using inflationary coordinates,
\beq \label{infla2} ds^{2}= -dr^{2} + e^{-2r/l_{\scriptscriptstyle
dS}}\left( dr^{2} + d{\bf x}^{2} \right), \eeq the spacelike
boundary of dS space, $I^{-}$, is located at $r=-\infty$. In this
paper we investigated solutions interpolating between asymptotic
regions of the form eq.~(\ref{poin2}) and (\ref{infla2}) and their
$r\rightarrow -r$ counterparts.

A more general formulation of the AdS/CFT and the dS/CFT dualities
considers an equivalence between asymptotically (A)dS spacetimes
and non-conformal field theories. In the context of the AdS/CFT
correspondence, this leads to the so-called UV/IR
correspondence\cite{UVIR,UVIR2} stating that asymptotic regions
(near the boundary) of AdS space are associated with short
distance (UV) physics in the field theory dual while regions
deeper in the AdS are related to long distance (IR) physics. It is
then natural to associate evolution along the $r$-coordinate with
a field theory renormalization group (RG) flow. An interesting
result then is the derivation of a gravitational
c-theorem\cite{ctheo}. The corresponding c-function is a local
geometric quantity which gives a measure of the number of degrees
of freedom relevant for physics in the dual field theory at
different energy scales. Requiring gravitational energy conditions
to be satisfied indicates that this c-function must decrease in
evolving from the UV toward the IR regions, as expected. Ideas
along the same lines can be used in the context of the dS/CFT
correspondence. Again demanding energy conditions to hold leads to
a gravitational c-theorem for asymptotically dS
spacetimes\cite{AS,BBM,tale1}. The latter states that the
c-function must decrease (increase) in a period of contraction
(expansion). For example, when considering spacetimes with both
asymptotia (at $r=\pm \infty$) of the form (\ref{infla2}), the
boundary $I^{-}$ corresponds to the UV while the horizon at
$r=+\infty$ is interpreted as an IR limit of the field theory.

Let us begin by considering spacetimes with two boundaries located
at $r= \pm \infty$. These spacetimes could be (A)dS/(A)dS or the
mixed geometries dS/AdS introduced in this paper. Following
ref.~\cite{tale1} we define an effective cosmological constant
\beq \label{lg} \Lambda_{\scriptstyle eff} =
-G_{\mu\nu}n^{\mu}n^{\nu}, \eeq where $n^{\mu}$ is unit vector
normal to the constant $r$ hypersurfaces. Expression (\ref{lg}) in
fact reduces to eqs.~(\ref{lads}) and (\ref{lds}) when evaluated
respectively for the spacetimes (\ref{poin2}) and (\ref{infla2}).
A reasonable definition for the c-function is \beq c\sim
\frac{1}{G_{\scriptscriptstyle N}|\Lambda_{\scriptstyle
eff}^{(d-2)/2}|}= \frac{1}{G_{\scriptscriptstyle
N}}|-G_{\mu\nu}n^{\mu}n^{\nu}|^{-(d-2)/2}. \eeq Of course,
introducing a gravitational c-function is relevant only if the
gravitational background under consideration has a field theory
dual. In the case of spacetimes interpolating between regions with
dS and AdS boundaries, it would be interesting to find out what
kind of field theory, if any, could be the dual. While we cannot
address this question to a satisfactory level at this point, one
thing we can do is describe roughly the rather strange field
theory RG flow such a peculiar gravitational background would be
dual to. The existence of such a mixed (dS/AdS)/CFT correspondence
would imply that the bulk coordinate which is reconstructed by the
field theory RG flow is timelike in the region containing the dS
boundary and spacelike in the region containing the AdS boundary.
Let us for the moment assume that the weak energy condition, which
was used in deriving the c-theorems associated to AdS/CFT and
dS/CFT correspondences, holds everywhere in the corresponding
gravitational dual. We would then expect the following to hold:
The boundary at $r=+\infty$ (AdS) corresponds to physics in the
conformal UV limit of the field theory. The weak energy condition
being satisfied for $r>0$, we expect that translations along
decreasing $r$, \ie  moving deeper inside of the spacetime,
corresponds to flowing toward an IR region close to $r=0$. Past
the horizon, the variable $r$ becomes timelike which means that
the RG flow should now describe a limit of the field theory which
is Euclidean. Since the weak energy condition is still assumed to
hold for $r<0$, the c-theorem guarentees that evolving toward
increasingly negative values of $r$ corresponds to `integrating
in' degrees of freedom, \ie moving toward another UV fixed point,
this time defined on the spacelike boundary $I^{-}$. Since both
these UV fixed points appear to be flowing toward the same IR
limit, it would be tempting to postulate a new self-duality for
the field theory. The latter would be a rather strange duality
between a Lorentzian and a Euclidean limit of the same field
theory. In this paper we found evidence that such spacetimes do
not exist. We have found that dS/AdS gravitational backgrounds can
exist only when energy conditions are violated in which case the
c-theorems are not valid anymore. Because they contain fields with
kinetic terms of the wrong sign, the type II* theories may be a
good place where dS/AdS geometries and consequently their field
theory dual can be studied.

\acknowledgments The authors would like to thank Jim Cline, Damien
Easson, Don Marolf, Eric Poisson and especially Rob Myers for
stimulating conversations. FL wishes to thank Sadik Deger for its
involvement at an early stage of this project. FL thanks the Isaac
Newton Institute for Mathematical Sciences for their hospitality
in the initial stages of this project. FL would also like to thank
the Perimeter Institute and the University of Waterloo's
Department of Physics for their ongoing hospitality. We were
supported in part by NSERC of Canada and Fonds FCAR du Qu\'ebec.
Finally, we would like to thank Jordan Hovdebo for proof-reading
an earlier draft of this paper.

\appendix
\section{Energy conditions and equation of state for scalars}
In this appendix we comment on the energy conditions and the
equation of state associated with a real scalar field in a
gravitational background of the form (\ref{metricgen}). The
stress-energy tensor components for a real scalar field are \beq
T^{r}_{r} = -\frac{(d-1)(d-2)}{2}\left[ -h\phi^{'2} + V(\phi)
\right], \eeq \beq T^{t}_{t} = T_{x}^{x} =
-\frac{(d-1)(d-2)}{2}\left[ h\phi^{'2} + V(\phi) \right]. \eeq For
$r>0$, \ie in the region containing the AdS boundary, the
components of the stress-energy tensor are \beq \rho = -T_{t}^{t}
= \frac{(d-1)(d-2)}{2}\left[ h\phi^{'2} + V(\phi) \right], \eeq
\beq p_{x}= T_{x}^{x} = -\frac{(d-1)(d-2)}{2}\left[ h\phi^{'2} +
V(\phi) \right],  \eeq and \beq p_{r}= T_{r}^{r} =
-\frac{(d-1)(d-2)}{2}\left[ -h\phi^{'2} + V(\phi) \right],\eeq
where $\rho $ is the conserved energy density associated with the
Killing direction $t$ and $p_{r}$, $p_{x}$ are the principal
pressures respectively along the directions $r$ and $x^{a}$. For
$r>0$ we have an anisotropic fluid since $p_{r}\neq p_{x}$. There
are two equations of state: \beq \frac{p_{x}}{\rho}=-1, \eeq and
\beq \frac{p_{r}}{\rho} =
\frac{h\phi^{'2}-V(\phi)}{h\phi^{'2}+V(\phi)}. \eeq Pure AdS and
dS space have $p/\rho=-1$ where $p$ represents all principal
pressures. In our case, the quantity $h\phi^{2'}+V(\phi)$ is
conserved but not $h\phi^{2'}-V(\phi)$ which means that their
ratio $p_{r}/\rho$ depends on the variable $r$. In order to have
an asymptotic $r\rightarrow +\infty$ region corresponding to AdS
we must therefore have \beq \label{asymptote3} \lim_{r\rightarrow
+\infty} \frac{h\phi^{'2}}{V(\phi)} = 0.\eeq In the region
containing the dS boundary ($r<0$) we have \beq \rho = -T_{r}^{r}
= \frac{(d-1)(d-2)}{2}\left[ -h\phi^{'2} +V(\phi) \right], \eeq
\beq p=p_{t} = p_{x} = T^{t}_{t} = -\frac{(d-1)(d-2)}{2}\left[
h\phi^{'2} + V(\phi) \right], \eeq which corresponds to the
equation of state \beq \frac{p}{\rho} =
\frac{h\phi^{'2}+V(\phi)}{h\phi^{'2}-V(\phi)}, \eeq which again
goes to $p/\rho=-1$ for large negative $r$ when
eq.~(\ref{asymptote3}) is satisfied. We note that for $r<0$ the
energy density $\rho $ is not conserved since there is no Killing
vector associated with the direction $r$. This is of course
expected for a time dependent background.

For $r<0$ we have \beq \rho + p_{x} = 0, \eeq and $\rho + p_{r} =
(d-1)(d-2)h(r)\phi^{'2}.$ Since $h(r)>0$ for $r>0$ we have
$\rho+p\geq 0$ in that region. Now for $r<0$ we find \beq \rho + p
= -(d-1)(d-2)h\phi^{'2},\eeq which is such that $\rho+p\geq 0$
since $h(r)<0$ for $r<0$. It is therefore easy to see that scalar
fields do not violate any of the energy conditions.


\begin{thebibliography}{99}

\bibitem{hull1}
C.~M.~Hull, ``Timelike T-duality, de Sitter space, large N gauge
theories and  topological field theory,'' JHEP {\bf 9807}, 021
(1998), hep-th/9806146.

\bibitem{hull2}
C.~M.~Hull, ``de Sitter space in supergravity and M theory,'' JHEP
{\bf 0111}, 012 (2001), hep-th/0109213.

\bibitem{AS1} A.~Strominger,
``The dS/CFT correspondence,''
JHEP {\bf 0110}, 034 (2001), hep-th/0106113].

\bibitem{AS} A.~Strominger,
``Inflation and the dS/CFT correspondence,''
JHEP {\bf 0111}, 049 (2001), hep-th/0110087.

\bibitem{BBM} V. Balasubramanian, J. de Boer, and D. Minic,
``Mass, Entropy, and Holography in Asymptotically de Sitter spaces''
hep-th/0110108].

\bibitem{witten} E.~Witten,
``Quantum gravity in de Sitter space, hep-th/0106109.

\bibitem{adsrev} O.~Aharony, S.S.~Gubser, J.~Maldacena, H.~Ooguri and
Y.~Oz, ``Large N field theories, string theory and gravity,''
Phys.\ Rept.\  {\bf 323},183 (2000), hep-th/9905111.

\bibitem{AS3}
M.~Spradlin, A.~Strominger and A.~Volovich, ``Les Houches lectures
on de Sitter space,'' arXiv:hep-th/0110007.

\bibitem{CF}
J.~M.~Cline and H.~Firouzjahi, ``No-go theorem for
horizon-shielded self-tuning singularities,'' Phys.\ Rev.\ D {\bf
65}, 043501 (2002) [arXiv:hep-th/0107198].

\bibitem{malik}
M.~K.~Parikh, I.~Savonije and E.~Verlinde, ``Elliptic de Sitter
space: dS/Z(2),'' arXiv:hep-th/0209120.

\bibitem{gubser}
O.~DeWolfe, D.~Z.~Freedman, S.~S.~Gubser and A.~Karch, ``Modeling
the fifth dimension with scalars and gravity,'' Phys.\ Rev.\ D
{\bf 62}, 046008 (2000) [arXiv:hep-th/9909134].

\bibitem{tale1}
F.~Leblond, D.~Marolf and R.~C.~Myers, ``Tall tales from de Sitter
space. I: Renormalization group flows,'' JHEP {\bf 0206}, 052
(2002), hep-th/0202094.

\bibitem{marolf}
D.~Marolf and S.~F.~Ross, ``Stringy negative-tension branes and
the second law of thermodynamics,'' JHEP {\bf 0204}, 008 (2002)
[arXiv:hep-th/0202091].

\bibitem{UVIR}
L.~Susskind and E.~Witten, ``The holographic bound in anti-de
Sitter space,'' hep-th/9805114.

\bibitem{UVIR2} A.W.~Peet and J.~Polchinski, ``UV/IR relations in AdS
dynamics,'' Phys.\ Rev.\ D {\bf 59}, 065011 (1999),
hep-th/9809022.

\bibitem{ctheo} D.Z.~Freedman, S.S.~Gubser, K.~Pilch and N.P.~Warner,
``Renormalization group flows from holography supersymmetry and a
Adv.\ Theor.\ Math.\ Phys.\  {\bf 3}, 363 (1999), hep-th/9904017;\\
J.~de Boer, E.~Verlinde and H.~Verlinde, ``On the holographic
renormalization group,''
JHEP {\bf 0008}, 003 (2000), hep-th/9912012;\\
V.~Balasubramanian, E.G.~Gimon and D.~Minic, ``Consistency
conditions for holographic duality,''
JHEP {\bf 0005}, 014 (2000), hep-th/0003147;\\
V.~Sahakian, ``Holography, a covariant c-function and the geometry
of the renormalization group,''
Phys.\ Rev.\ D {\bf 62}, 126011 (2000), hep-th/9910099;\\
E.~Alvarez and C.~Gomez, ``Geometric holography, the
renormalization group and the c-theorem,'' Nucl.\ Phys.\ B {\bf
541}, 441 (1999), hep-th/9807226.

\end{thebibliography}
\end{document}